# BUILDING SUSTAINABLE
# ECOSYSTEM-ORIENTED ARCHITECTURES

Youssef Bassil

LACSC – Lebanese Association for Computational Sciences
Registered under No. 957, 2011, Beirut, Lebanon
youssef.bassil@lacsc.org

## ABSTRACT

*Currently, organizations are transforming their business processes into e-services and service-oriented architectures to improve coordination across sales, marketing, and partner channels, to build flexible and scalable systems, and to reduce integration-related maintenance and development costs. However, this new paradigm is still fragile and lacks many features crucial for building sustainable and progressive computing infrastructures able to rapidly respond and adapt to the always-changing market and environmental business. This paper proposes a novel framework for building sustainable Ecosystem-Oriented Architectures (EOA) using e-service models. The backbone of this framework is an ecosystem layer comprising several computing units whose aim is to deliver universal interoperability, transparent communication, automated management, self-integration, self-adaptation, and security to all the interconnected services, components, and devices in the ecosystem. Overall, the proposed model seeks to deliver a comprehensive and a generic sustainable business IT model for developing agile e-enterprises that are constantly up to new business constraints, trends, and requirements. Future research can improve upon the proposed model so much so that it supports computational intelligence to help in decision making and problem solving.*

## KEYWORDS

*Ecosystem Oriented Architecture, Digital Ecosystem, Service Science, Sustainable Computing*

## 1. INTRODUCTION

At present, enterprises are increasingly focusing on transforming their traditional core businesses into e-services to create an agile distributed e-business [1]. Typically, electronic business allows a seamless interaction between a company and its clients, as well as its partners and associates. On the long run, this improves its productivity, increases its profitability, and reinforces its market power. A way for building e-service models is SOA short for Service-Oriented Architecture. Fundamentally, SOA is a model for system development based on loosely-integrated suite of services that can be used within multiple business domains [2, 3]. In practice, SOA has many benefits: it promotes the reusability of existing technological assets; it accelerates the expandability and evolution of information systems; it eases systems integration; and it simplifies the building of next-generation composite applications [4, 5, 6]. Thus, it reduces application development and maintenance costs, improves coordination across the different business stakeholders and processes, and increases business agility to respond quickly to on-demand requirements. Although SOA had great success [7], it has many limitations and drawbacks [8, 9], and lacks many features such as universal interoperability i.e. the ability to mesh incompatible and wide-ranging technologies; manageability i.e. the ability to be autonomously operated; adaptability i.e. the ability to self-adapt according to the state of its resources and execution environment; integrability i.e. the ability to autonomously discover and self-integrate new service components; survivability i.e. the ability to survive a disaster;

availability i.e. the ability to stay up without any downtime; and security i.e. the ability to defend and protect against malware threats [10]. Altogether, these aforesaid characteristics, if provided, would deliver a development model for building sustainable service-based information systems. This paper presents a generic framework for building the successor of SOA, namely EOA short for Ecosystem-Oriented Architecture. EOA defines a digital ecosystem with properties of sustainability, universal interoperability, manageability, self-integration, self-adaptation, and security [10], inspired by natural ecosystems for building business models and architectures for sophisticated, distributed, and collaborative e-enterprises, e-marketplaces, e-communities, and e-cities using reusable service components [11]. The backbone of this framework is an ecosystem layer comprising several computing units: Ecosystem Management Bus (EMB), Ecosystem Communication Unit (ECU), Ecosystem Integration Unit (EIU), Ecosystem Management Language (EML), Ecosystem WMI Scripting Unit (EWSU), and Ecosystem Security Unit (ESU). Their aim is to provide standardization, transparent communication, automated management, self-integration, self-adaptation, and security for all the interconnected hardware and software in the digital ecosystem.

## 2. SOA LIMITATIONS

Traditional service-oriented and component-based architectures such as SOA provide computational resources as loosely and distributed components called services. However, these models do not exhibit sustainability features including universal interoperability, manageability, self-integration, self-adaptation, and security [12]. As a result, new challenges have come to light and can be summarized as follows:

Universal Interoperability: SOA does not allow a standardized and an effective interaction and data exchange between a wide range of products, manufactured by different vendors and service providers, and built using different technologies and platforms.

Manageability: SOA does not define protocols and high-level languages to manage and control its components in a consistent, efficient, and automated manner. A manageable system is a system that has the ability to be corrected, updated, expanded, and partially replaced without impacting other components in the system.

Self-Integration: SOA does not provide a mechanism for the automatic discovery and integration of components and services into the existing infrastructure. For instance, an SOA cannot be scaled nor have its modules replaced without requiring programs to be modified and recompiled.

Self-Adaptability: SOA does not provide a mechanism to self-optimize itself and adapt its internal state according to the state of its execution environment. This includes increasing automatically disk storage to cope with data growth, applying load balancing techniques to contain the increase of users, assigning extra processing cycles to computationally intense applications, and requesting more Internet bandwidth for rich-Internet applications.

Security: The decentralized nature of SOA is by itself a threat. Since all services are not located locally as in personal computers, they are subject to security breaches and vulnerabilities. Moreover, SOA's environment is not encrypted and therefore encrypting messages, deploying firewalls, and forcing role-based policies are crucial to ensure the self-protection of the system.

Sustainability: SOA does not feature all the above-mentioned attributes. As a result, it is inadequate to build information systems that are cross-platform, automatically manageable, adaptably optimizable under severe circumstances, easily scalable, secure, reliable, and can stay active for a long period of time, while withstanding system failures, stoppages, and bugs.

## 3. RELATED WORK

Over years, several studies have been done to enhance SOA architectures and provide significant improvements for their functionalities, features, and implementations. In this section, several of these attempts are to be discussed elaborately.

### 3.1. CAWE Framework

The composition model proposed in SOA does not explicitly deal with personalization and context-awareness. In order to address such limitations, the CAWE (Context Aware Workflow Execution) conceptual framework was conceived [13]. CAWE is used to develop context-aware composite web applications for SOA architectures with such properties as the support of execution of context-sensitive workflows, the capability to easily manage user interactions, and the ability to personalize information based on different users and devices. All in all, CAWE adds self-adaptation properties to service-oriented architectures allowing them to meet the requirements of heterogeneous users in an always-changing environment.

### 3.2. Self-Integration

A major challenge in SOA is the integration of services into the existing infrastructure which is so far a non-automated process. For this reason, a model for self-integrating services in SOA architectures was proposed [14]. It is based on WSDL documents which allow the automation of web service discovery, integration, deployment, and monitoring processes. Below are the system's original steps:

Service discovery: The WSDL documents which represent service queries are matched against the service offers. A matching score is then calculated and used to rank the different services.

Service integration: The discovery algorithm produces a collection of mediator plug-ins for successful matches. These mediators are used at runtime to enable ad-hoc service integration.

Deployment process: Mediators are deployed in the system and their endpoints are invoked by the participating web services.

Monitoring: All unreachable services are removed and the system will start a new round to discover new services.

### 3.3. Survivability

Survivability is the ability of a system to continue to operate in spite of errors, failures, or accidents [15]. In fact, SOA does not define a method for error recovery, nor a clear scheme for building survivable components. In view of that, an error recovery model was proposed [16]. It contains four different states: good state, vulnerable state, fault state, and recovery state. Initially, the system is in the good state implying that it is operating correctly. It moves to the vulnerable state whenever a user violates a security policy, for instance, accessing a resource without authorization. It then enters the fault state when vulnerability is successfully exploited. As a result, the system automatically transits to the last state, namely the recovery state, wherein the system is totally recovered. Figure 1 depicts the state transition diagram for this proposed error recovery model.

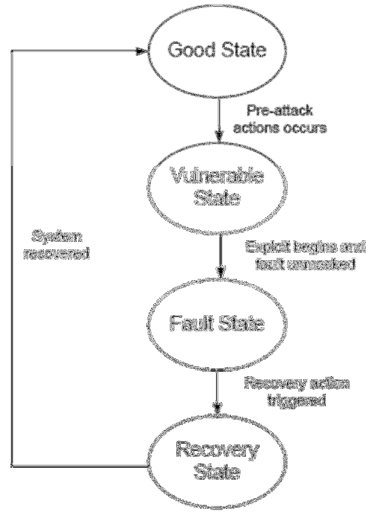

Figure 1. Error recovery state transition diagram

## 3.4 Sustainability

The OASIS reference model [17] is a generic framework for building and managing service-oriented architectures. It is majorly composed of six units: the orchestration and management unit which is responsible for administering the connected components and web services in the SOA; the data content unit which represents a set of databases that feed web services with data and information; the service description unit which defines the functions exposed by the connected web services in the SOA; the service discovery unit which contains a look-up registry to locate and consume web services; the messaging unit which can be thought as the communication medium that lets all connected components share data and communicate between each other; and the security and access unit which provides a security layer for securing and encrypting the messages being sent and received between the different components of the SOA. Figure 2 depicts the OASIS reference model.

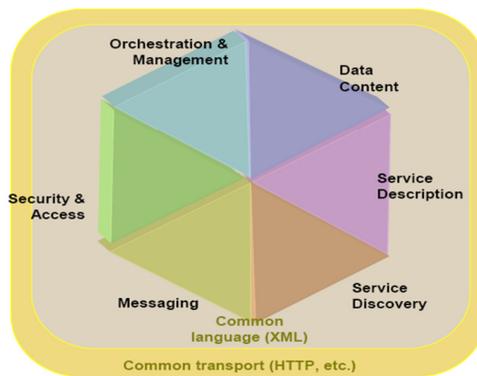

Figure 2. OASIS model

## 4. ECOSYSTEM-ORIENTED ARCHITECTURE

Unlike the service-oriented architecture which only contains three basic layers, mainly the presentation, the service, and the data layer, the proposed ecosystem-oriented architecture (EOA) adds an additional layer called the ecosystem layer whose role is to deliver a sustainable operational environment for the underlying IT infrastructure. This includes such features as data-path and messaging middleware, universal interoperability, transparent communication, automated management, self-integration, self-adaptation, and security for all services in the ecosystem. Figure 3 depicts the four layers of the proposed EOA.

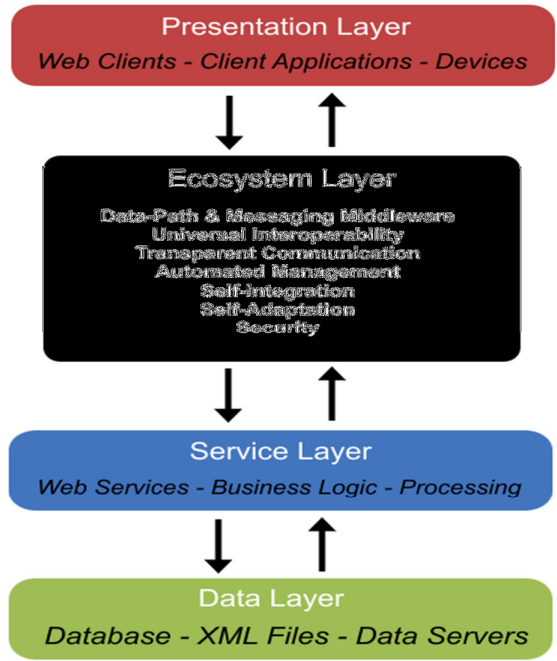

Figure 3. The four layers of the proposed EOA

## 4.1. The Presentation Layer

The presentation layer is the top-most level of functionalities that mainly provides such system's input and output interfaces as sending requests to and receiving responses from services, browsing catalog, buying merchandise, invoking remote content, and reporting.

## 4.2. The Service Layer

The service layer defines the execution of the application, processes clients' requests, makes logical decisions and evaluations, and performs intensive calculations. Usually, web services, programming modules, dynamic libraries, and APIs are deployed in this layer.

## 4.3. The Data Layer

The data layer is where data are stored and retrieved. Typically, data are saved into databases, system files, or even XML files. The service layer requests data for processing from the data layer and then passes the results back to the presentation layer.

## 4.4. The Proposed Ecosystem Layer

In effect, the proposed ecosystem layer is a middleware sitting between the presentation and the service layer and providing several functionalities and features. It is made out of six building blocks or operational units. Figure 4 depicts these six units along with their tasks and characteristics.

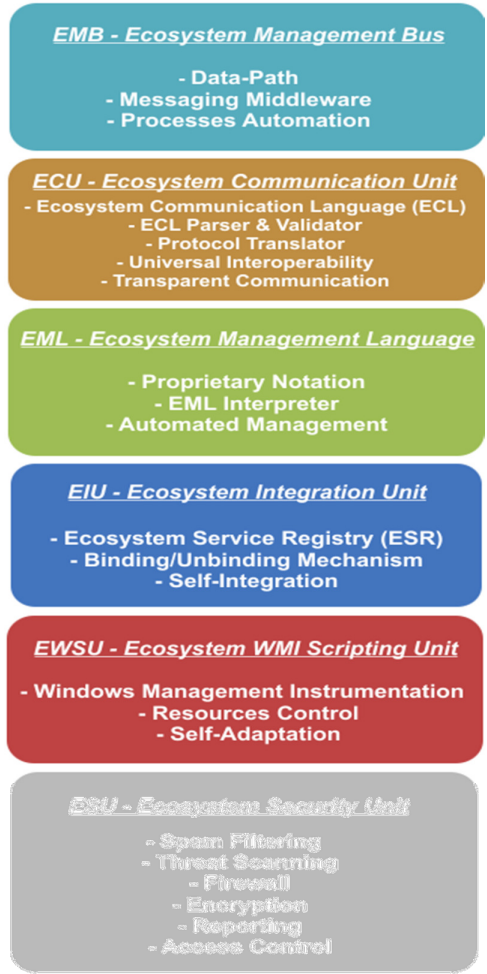

Figure 4.  The six units of the Ecosystem layer

The Ecosystem Management Bus (EMB) delivers the data-path of the entire ecosystem and a messaging middleware for sending and receiving messages between the different interconnected services.

The Ecosystem Communication Unit (ECU) delivers standardization and a transparent communication protocol defined by a proprietary XML-based language that allows the collaboration and the interoperability of different services built using different architectures, programming languages, and technologies.

The Ecosystem Management Language (EML) delivers automated management for services using a proprietary high-level language based on proprietary syntax rules and vocabulary. The prime role of EMB is to manage, control, monitor, and administer the distributed services inside the ecosystem.

The Ecosystem Integration Unit (EIU) delivers self-integration by automating service discovery, integration, and deployment.

The Ecosystem WMI Scripting Unit (EWSU) delivers self-adaptation by allowing the running ecosystem to optimize itself and change its state according to the changes in the execution environment.

The Ecosystem Security Unit (ESU) delivers security technologies to protect information traveling through the ecosystem such as message encryption, spam filtering, and firewall.

### 4.4.1. EMB - Ecosystem Management Bus

The Ecosystem Management Bus (EMB) provides a data-path for data to travel between the functional units and services of the ecosystem. It constitutes a data transmission medium, emulating a messaging middleware that bridges between the different interconnected and distributed services to allow them send and receive data back and forth to each other. Characteristically, it automates the in and out communications between all involved parties and coordinates the interaction between them, and allows the storage, routing, and transformation of messages during inter-system interactions. Furthermore, the EMB houses the different functional units of the ecosystem and thus it works as a central orchestrator that manages the operations going inside the ecosystem. Figure 5 shows the actual diagram of the EMB.

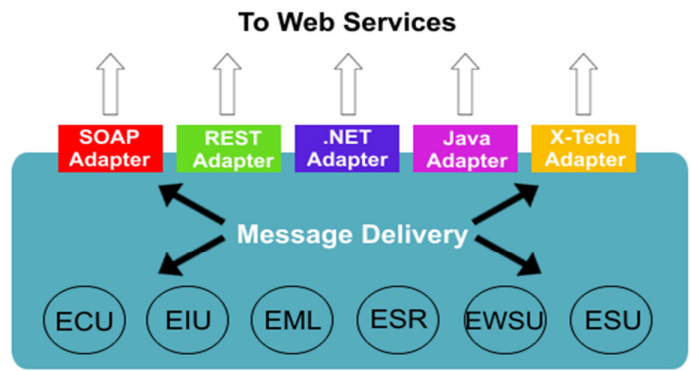

Figure 5.  EMB

### 4.4.2. ECU - Ecosystem Communication Unit

The Ecosystem Communication Unit (ECU) is a multi-agent and a multi-platform design for connecting services, possibly incompatible, together to interact, send requests, and receive responses from each other. As a multi-agent model, it permits the distribution of services over different machines, networks, and premises, allowing a seamless and transparent communication between them. As a multi-platform model, it permits the support of incompatible services built using different platforms, different standards, different technologies, and different programming languages. Additionally, the ECU provides a communication language called ECL short for Ecosystem Communication Language based on XML language for exchanging structured information between the different services of the ecosystem. ECL is based on XML syntax to format messages sent to and received from inner-system services. In practice, a client requesting an operation sends an ECL message with the appropriate parameters to a destination service. The service returns then an XML-formatted response with the resulting data. Being based on standard message format, ECL promotes interoperability and standardization for all services regardless of their implementation, target-platform, and underlying technology. Following is a sample request in ECL language. It consists of a sender application whose IP is 192.168.1.20 and ID is 24 invoking a function called "Max" with two integer parameters 10 and 50 respectively, over a service whose IP is 192.168.1.177 and ID is 91

```
<protocol>
    <sourceIP>192.168.1.20</sourceIP>
    <destinationIP>192.168.1.177</destinationIP>
    <sourceID>24</sourceID>
    <destinationID>91</destinationID>
    <functionInvoked>Max</functionInvoked>
    <functionParams>
        <param>10</param>
        <type>int</type>
        <param>50</param>
        <type>int</type>
    </functionParams>
    <functionReturnType>int</functionReturnType>
    <stamp>5/4/2011 09:32:10PM</stamp>
    <version>1.0</version>
</protocol>
```

### 4.4.3. EML - Ecosystem Management Language

The Ecosystem Management Language (EML) is a declarative language based on a proprietary syntax used to administer every single component connected to the ecosystem infrastructure. At heart, its purpose is to ease and automate the management and control of the ecosystem using control commands issued by administrators via a console manager. For instance, one of EML's commands is the "bind" command which is used to connect a new web service into the system, while "unbind" is used to disconnect it. The "is-run" command is used to check whether or not an existing service is in online or offline mode. Another command can grant and revoke security permissions from a specific service; whereas, the command "replica" creates a replication for an existing service.

The core of the EML is an EML interpreter which scans an issued EML command, extracts valuable tokens out if it, parses them to validate their correct arrangement, and then executes the command. Generally speaking, EML helps better automate the management and administration of the different operating services in the ecosystem.

### 4.4.4. EIU - Ecosystem Integration Unit

The Ecosystem Integration Unit (EIU) facilitates the discovery, self-integration, and dis-integration of services in and out of the existing ecosystem. The process starts when a new service needs to integrate into the present infrastructure. The EIU intervenes to validate the Service Description Language (SDL) of the service that is requesting integration. If validation is successful, an acknowledgement is sent to the corresponding service and a new record is created in the Ecosystem Service Registry (ESR) containing important details such as service ID, service protocol, service IP, service functions, parameters, and return data type. Figure 6 depicts the EIU and the various steps required to self-integrate a new service into the ecosystem.

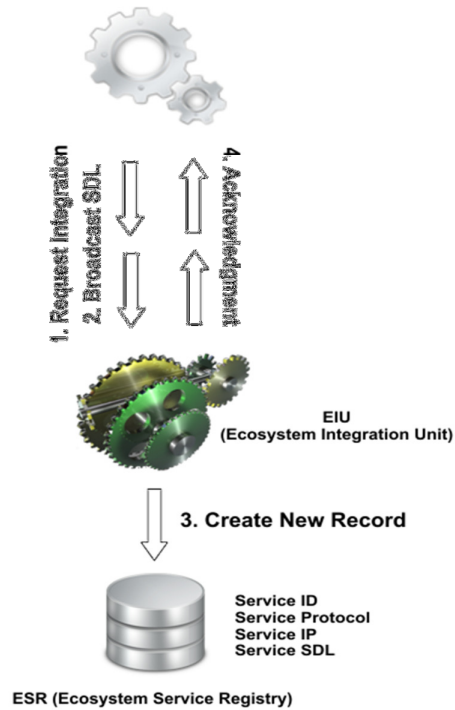

Figure 6.  EIU

The algorithm can be summarized as follows:

**ALGORITHM**

// send a request for integration

// broadcast SDL

**if**(validation(params)==true)

{

InsertRecord(randomID, protocol, serviceIP, SDL) // insert details into ESR

// this ESR record contains a unique random ID that uniquely identifies a service in the system,

// the service protocol either SOAP, REST, or any other technology,

// the service IP which indicates the Internet address of the machine hosting this new service,

// SDL contains the list of functions that the new service encapsulates along with their parameters and data types.

// send positive acknowledgment message

}

**else**

// reject request

### 4.4.5. EWSU - Ecosystem WMI Scripting Unit

The role of the Ecosystem WMI Scripting Unit (EWSU) is to provide self-adaptation for the ecosystem allowing it to change its state based on the state of its execution environment such as increasing memory allocation, increasing disk quota, assigning more CPU cores and cycles, and reducing power consumption. For this reason, EWSU utilizes the Windows Management Instrumentation (WMI) API [18] which uses scripts to automate administrative tasks on remote computers as well as the management of the operating system peripherals, resources, and products. Fundamentally, WMI is a set of extensions to the Windows Driver Model to monitor product's performance, diagnose errors, write trace information, manage operating system's resources, and provide system information and notification. In order to deliver instrumentation for ecosystem-oriented architectures, WMI is incorporated inside the ecosystem layer. Upon the execution of a specific script, the EWSU engine interprets it and hands it to the WMI COM API which passes it, in turn, down to the Windows WMI Framework. This framework will then execute it over the corresponding driver which will accordingly change the behavior and the settings of the actual hardware. Below is the general syntax for executing a WMI script using EWSU.

*executeWMI: Service-ID, WMI-script*

*executeWMI-ack: Service-ID, True\False*

### 4.4.6. ESU - Ecosystem Security Unit

The Ecosystem Security Unit (ESU) provides all sort of protection against malwares and attacks, and ensures the correct implementation of security polices and access controls inside the ecosystem. Essentially, the ESU provides several security technologies that are listed below:

Spam Filtering [19]: It isolates unsolicited requests to the services of the ecosystem. Besides, it monitors and inspects every single message that circulates throughout the ecosystem based on its content, size, origin, and type.

Threat Scanning: It captures and quarantines viruses, spywares, trojans, and backdoors [20] so as to ensure maximum protection while the ecosystem is running. It also detects and prevents several security attacks and threats such as DoS (Denial of Service), IP spoofing, session hijacking, DNS poisoning, and password cracking.

Firewalls [21]: They block unwanted ports and Internet addresses from network transmission. Further, they inspect each packet passing through the network and accept or reject it based on source and destination port, source and destination IP, service ID, and other user-defined rules.

Encryption [22]: It ciphers all messages that travel in and out of the ecosystem. The ESU provides several encryption algorithms to guarantee data concealment all the time. For instance, AES (Advanced Encryption Standard) is used to encrypt the communication between all the components of the ecosystem; while, SSL is used to securely encrypt HTTP web requests sent between the different services.

Access Control: It grants and revokes permissions based on users and service identities. The ESU employs an Access Control Matrix [23] that represents the rights of each subject with respect to every object in the system. Subjects are the entities that can perform actions, while objects are the resources on which access needs to be controlled.

Reporting and Logging: They log all communication and activities during the routine operation of the ecosystem. They have also the ability to capture detailed history for the invoked services including their IDs, IPs, timestamps, functions requested, protocols used, bytes served, and user-agents. In addition, the ESU can report errors, runtime exceptions, and any kind of faults that have occurred during the normal operation of the ecosystem. Figure 7 displays the six major security technologies provided by the ESU.

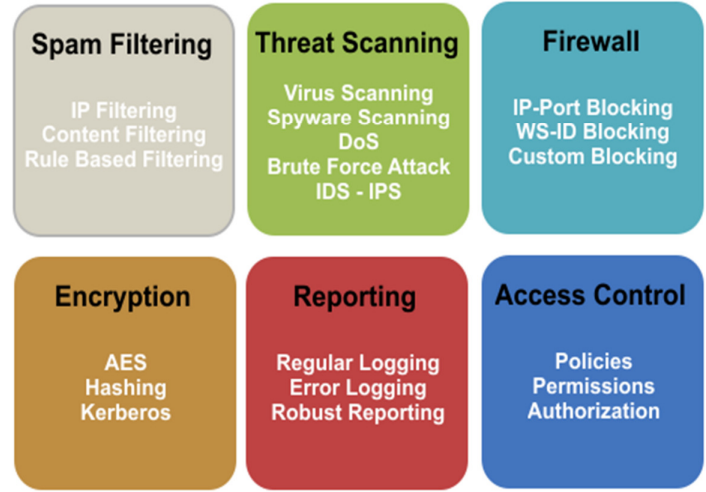

Figure 7 – ESU

## 5. CONCLUSIONS

This paper presented a conceptual model for building sustainable ecosystem-oriented architectures with such features as universal interoperability, manageability, self-integration, self-adaptation, and security. The proposed model is a collection of several units, each having a particular purpose and function. The EMB (Ecosystem Management Bus) is the central data-path responsible for delivering messages between different services. The ECU (Ecosystem Communication Unit) features a communication language called ECL to format requests and responses of collaborating services. The EML (Ecosystem Management Language) is a proprietary language made out of high-level commands to automate the administration of the inner services of the ecosystem. The EIU (Ecosystem Integration Unit) is responsible for binding and unbinding services in a consistent and automated manner. The EWSU (Ecosystem WMI Scripting Unit) is responsible for self-adapting the ecosystem and allocating and de-allocating resources based on the requirements and needs. The ESU (Ecosystem Security Unit) provides a secure operating environment for working services and an ultra-tight protection for the entire ecosystem against malicious network attacks. This complete framework allows the building of sustainable and avant-garde large-scale computing service-based models that leverage existing technological assets, reduce application development costs, and promote the development of agile e-enterprises that can cope with the ever-changing e-demands, trends, and business requirements.

## 6. FUTURE WORK

The proposed sustainable ecosystem-oriented architecture can be improved in several ways, one of which is adding computational intelligence to the ecosystem layer such as knowledge storage to give the system the ability not only to process raw data but also to infer, reason, and help in decision making and problem solving. Above and beyond, the EML and the ECL languages could be extended to provide richer functionalities allowing more control over the different interconnected services of the ecosystem.

## ACKNOWLEDGEMENTS

This research was funded by the Lebanese Association for Computational Sciences (LACSC), Beirut, Lebanon under the "Digital Ecosystem Research Project – DERP2011".